\documentclass[aps,twocolumn,nofootinbib,superscriptaddress,prd]{revtex4-1}
\usepackage{amssymb}
\usepackage{amsmath}
\usepackage{amstext}
\usepackage{amsfonts}
\usepackage{bbold}
\usepackage{bm}
\usepackage{graphics}
\usepackage{mathtools}
\usepackage{lmodern}
\usepackage{graphicx}
\usepackage[capitalise]{cleveref}
\usepackage{xcolor}
\usepackage[export]{adjustbox}

\def\ba{\begin{align}}
\def\ea{\end{align}}

\begin{document}

\title{Sub-GeV Dark Matter in Superfluid He-4: an Effective Theory Approach}

\author{Francesca Acanfora}
\affiliation{Dipartimento di Fisica, Sapienza Universit\`a di Roma, P.le A Moro 2, Roma, I-00185, Italy}
\author{Angelo~Esposito}
\affiliation{Theoretical Particle Physics Laboratory (LPTP), Institute of Physics, EPFL, 1015 Lausanne, Switzerland}
\author{Antonio~D.~Polosa}
\affiliation{Dipartimento di Fisica and INFN, Sapienza Universit\`a di Roma, P.le Aldo Moro 2, I-00185 Roma, Italy}

\begin{abstract}
We employ an effective field theory to study the detectability of sub-GeV dark matter through its interaction with the gapless excitations of superfluid $^4$He. In a quantum field theory language, the possible interactions between the dark matter and the superfluid phonon are solely dictated by symmetry. We compute the rate for the emission of one and two phonons, and show that these two observables combined  allow for a large exclusion region for the dark matter masses. Our approach allows a direct calculation of the differential distributions, even though it is limited only to the region of softer phonon excitations, where the effective field theory is well defined. The method presented here is easily extendible to different models of dark matter.
\end{abstract}

\maketitle


\section{Introduction}

The existence of dark matter is one of the most compelling indications for physics beyond the Standard Model, and the question about its nature is hence of great interest. In recent years, following the negative results in the search of Weakly Interacting Massive Particles, more attention has been paid to the hypothesis of a dark matter with mass below the GeV, as suggested by different models --- see e.g~\cite{Boehm:2003hm,Boehm:2003ha,Hooper:2008im,Feng:2008ya,Zurek:2008qg,Hochberg:2014dra,Kuflik:2015isi,Kaplan:2009ag,Falkowski:2011xh,Hall:2009bx} and~\cite{Alexander:2016aln,Battaglieri:2017aum,Knapen:2017xzo} for recent reviews.

Given the very soft recoils expected, sub-GeV dark matter particles require new detection methods. 
Several ideas have been proposed in the literature, from  semiconductor targets~\cite{Essig:2011nj,Essig:2015cda,Graham:2012su} to superconductors~\cite{Hochberg:2015pha} and Fermi-degenerate materials~\cite{Hochberg:2015fth}.  Directionality in two-dimensional materials has been discussed in~\cite{Capparelli:2014lua, Cavoto:2016lqo, Hochberg:2016ntt, Cavoto:2017otc};  an intermediate program for the direct directional detection of MeV dark matter using graphene is planned  in the Ptolemy experiment~\cite{Baracchini:2018wwj}.

In this paper we will concentrate on the proposal to search  light invisible particles from scatterings in superfluid $^4$He targets, as  presented in~\cite{Schutz:2016tid,Knapen:2016cue}. The calorimetric readout of a superfluid $^4$He target is discussed in~\cite{Hertel:2018aal} and an account on particle detection by evaporation from superfluid helium can be found in~\cite{Hertel:2018aal,Bandler:1992zz,Maris:2017xvi}.

Indeed $^4$He offers several advantages such as a low target mass to maximize the energy deposited by the dark matter, high purity against radioactive decay and a suppressed background from electronic excitations. 

In~\cite{Schutz:2016tid,Knapen:2016cue} it has been proposed to look for a process where the dark matter interacts with the helium target, with consequent emission of an off-shell phonon (i.e. not sitting on the dispersion curve), which then decays into two on-shell ones. The expectation is that, although phase space suppressed, this process should maximise the energy released to final state phonons, potentially allowing for the detection thanks to an appreciable change in temperature of the superfluid.

In this work we employ an effective field theory (EFT) approach~\cite{Leutwyler:1996er,Son:2002zn,Nicolis:2011cs,Nicolis:2015sra} to describe the interaction between the dark matter and the superfluid phonon. The method we use, 
being solely based on symmetry arguments, is general to all superfluids (even the strongly coupled ones, like $^4$He) and allows to easily couple the dark matter to the phonon, using  standard quantum field theory methods. No approximate models  of the superfluid are required. The parameters of the effective theory are extracted from experiment.
We work in a relativistic setting, and take the nonrelativistic limit when appropriate.

We  re-evaluate the relevance of the emission of a single phonon.
When allowed by kinematics, this process is dominant and can offer one additional search channel, which is relevant for dark matter masses larger than 1~MeV. Moreover, the emission happens at \v{C}herenkov angles, which could allow to determine the direction of the incoming dark matter.

The plan of action for our analysis is the following. We write down the effective action, $S_\text{bulk}$, that describes the bulk of the superfluid alone, i.e. the phonon and its self-interactions. We then introduce the dark matter field and write the most general action, $S_\text{eff}$, for its coupling with the superfluid phonon. This action comes with effective coefficients that are a priori unknown. To estimate them, we consider a microscopic model, $S_\text{dark}$, for the dark matter particle and its interaction with $^4$He, which we match with the effective action above and use to estimate the effective couplings.

In this work, we consider a scalar dark matter charged under some dark $U_d(1)$ group, interacting with ordinary matter through a heavy mediator. 
The method illustrated can be extended to any model of dark matter, the only necessary input being the symmetries of the dark sector and its coupling to ordinary matter.

\vspace{1em}

\noindent\emph{Conventions:} Throughout this paper we set $\hbar=1$ and work with a metric signature $\eta_{\mu\nu}=\text{diag}(-1,1,1,1)$. In most of the paper we will also set $c=1$, except when explicitly stated.


\section{The EFT for superfluids}

The EFT approach to the description of gapless excitations in generic media describes the latter in terms of spontaneous symmetry breaking. Indeed, all media spontaneously break at least part of the Poincar\'e group. In particular, every condensed matter system breaks Lorentz boosts by singling out a particular reference frame: the one where the system is at rest. Other components of the group could be broken as well, and different symmetry breaking patterns characterise different states of matter~\cite{Nicolis:2015sra}. The associated Goldstone modes correspond to the collective excitations of the medium (see e.g.~\cite{Leutwyler:1996er,Son:2002zn,Endlich:2010hf,Nicolis:2011cs,Nicolis:2012vf,Endlich:2012pz,Hoyos:2013eha,Alberte:2018doe,Esposito:2018sdc}).

A zero-temperature s-wave superfluid is a system where a global $U(1)$ charge (particle number) is spontaneously broken by a background at finite density\footnote{The prototypical example is that of a gas of weakly coupled bosons. At zero temperature they all condense on the ground state, and the total wave function spontaneously breaks the particle number operator.}, where the vacuum expectation value (vev) of its generator, $\langle N\rangle$, is the number of particles~\cite{Nicolis:2011pv}. The ground state $|\mu\rangle$ of a finite density system is defined as the state that minimizes the modified Hamilotonian $\bar H = H-\mu N$, i.e. $\bar H |\mu\rangle = 0$. It then follows that, if such a state spontaneously breaks $N$, then time-translations must be broken as well, while the combination $\bar H$ remains unbroken.
On this background the energy of the system (i.e. the vev of the Hamiltonian) is $\langle H\rangle=\mu \langle N\rangle$.

Given the above symmetry breaking pattern, the simplest way to describe the low-energy dynamics of a superfluid is arguably in terms of a real scalar field\footnote{In the particular case of a weakly coupled gas of bosons, $\psi$ corresponds to the phase of the superfluid wave function~\cite{pethick2002bose,schakel2008boulevard}.} that shifts under the $U(1)$, $\psi\to\psi+\alpha$, and acquires a vev proportional to time\footnote{Even though the vev is divergent for large times, this has no consequences on observables, since the field $\psi$ always appears derived.}, $\langle\psi(x)\rangle=\mu t$. This background breaks boosts, time translations and the internal $U(1)$, but preserves the correct linear combination of the last two, as explained above. Nevertheless, the system admits a single Goldstone boson --- the superfluid phonon --- corresponding to the fluctuation of the field around equilibrium, $\psi(x)=\mu t+\pi(x)$.
Note that $\mu$ is the relativistic chemical potential, related to the more standard nonrelativistic one, $\mu_\text{nr}$, by $\mu=m+\mu_\text{nr}$, with $m$ the mass of the constituents of the superfluid\footnote{In fact, the chemical potential corresponds to the energy of the system per particle. In a relativistic framework, this gets a contribution from the particle's rest mass}.

Since the breaking of the above symmetries is spontaneous, the most general low-energy action for the scalar field must be invariant under the Poincar\'e group and the internal $U(1)$. At lowest order in the derivative expansion the only possibility is~\cite{greiter1989hydrodynamic,Son:2002zn}
\begin{align} \label{eq:Sbulk1}
S_\text{bulk}=\int d^4x\,P(X) \quad\text{ with }\quad X=\sqrt{-\partial_\mu\psi\partial^\mu\psi}\,,
\end{align}
where $P$ is a generic function.
Here $X$ is the local chemical potential, which differs from the background one in presence of fluctuations. The stress-energy tensor of this theory is
\begin{align}
T^{\mu\nu}=\eta^{\mu\nu}P(X)+P^\prime(X)\frac{\partial^\mu\psi\partial^\nu\psi}{X}\,,
\end{align}
where the prime denotes derivatives with respect to $X$ (or, equivalently, $\mu$).
From the above equation one deduces that $P(X)$ is the pressure of the superfluid.

Expanding the lagrangian up to cubic order in small fluctuations, one finds the action for the superfluid phonon
\begin{align} \label{eq:Sbulk}
\begin{split}
S_\text{bulk}&\supset \frac{\bar n}{\mu c_s^2}  \int  d^4x \bigg[\frac{1}{2}\dot\pi^2-\frac{c_s^2}{2}(\bm\nabla\pi)^2 \\
&\qquad\qquad\qquad\; +\lambda_3\dot\pi(\bm\nabla\pi)^2+\lambda_3^\prime\dot\pi^3\bigg] \\
&\to \int  d^4x \bigg[\frac{1}{2}\dot\pi^2-\frac{c_s^2}{2}(\bm\nabla\pi)^2 \\
&\qquad\qquad\,\; +\lambda_3\sqrt{\frac{\mu}{\bar n}}c_s\dot\pi(\bm\nabla\pi)^2+\lambda_3^\prime\sqrt{\frac{\mu}{\bar n}}c_s\dot\pi^3\bigg]\,,
\end{split}
\end{align}
where in the second line we have canonically normalized the field $(\pi\to\sqrt{\mu/\bar n}\,c_s\pi)$.
The sound speed $c_s$ and effective couplings are related to the pressure by
\begin{align}
c_s^2=\frac{P^\prime}{\mu P^{\prime\prime}}\,,\qquad\lambda_3=\frac{c_s^2-1}{2\mu}\,,\qquad\lambda_3^\prime=\frac{1}{6}\frac{\mu c_s^2}{\bar n}P^{\prime\prime\prime}\,,
\end{align}
where the derivatives are evaluated on the background, $X=\mu$. 
The background number density is  given by $\bar n=P^\prime$ (again by inspection of the stress-energy tensor). The only information necessary to extract all the effective parameters is the superfluid equation of state (e.g. $P=P(\mu)$ or $c_s=c_s(P)$)~\cite{abraham1970velocity}.
Finally, the propagator for a phonon with energy $\omega$ and momentum $\bm{q}$ reads
\begin{align}
G_\pi(\omega,\bm q)=\frac{i}{\omega^2-c_s^2\bm q^2+i\epsilon}\,.
\end{align}


\section{Dark matter-phonon interaction}

Let us now describe the interaction between the dark matter and the phonon. In our toy model the dark matter is described by a scalar field, $\chi(x)$, charged under some dark $U_d(1)$. We also assume that the dark sector is weakly coupled, and that its interaction to ordinary matter goes through a massive scalar mediator, $\phi(x)$.

Since we are interested in processes with one incoming and one outgoing dark matter particle, we look for the coupling between two dark matter fields and the superfluid phonon. The effective theory that describes such an interaction must be invariant under Poincar\'e transformations, the superfluid $U(1)$ and the dark $U_d(1)$. The most general low-energy effective action for the case of interest is then
\begin{align} \label{eq:Seff1}
\begin{split}
S_\text{eff}=&-\int d^4x\bigg[Z(X)|\partial\chi|^2+m^2(X)|\chi|^2  \\
&\qquad\qquad\;+A(X)\chi^\dagger\partial_\mu\chi  \partial^\mu\psi + \text{h.c.} \\
&\qquad\qquad\;+ B(X) \partial_\mu\chi\partial_\nu\chi^\dagger\partial^{\{\mu}\psi\partial^{\nu\}}\psi\bigg]\,,
\end{split}
\end{align}
where with $\{\,\dots\}$ we indicate the traceless combination of indices.
Note that any function of $X$ is invariant under the full symmetry group. Here $m^2(X)$ is the effective mass of the dark matter in medium, in analogy to the Archimedean principle. The action above contains all possible interactions between two dark matter fields and any number of phonons, at lowest order in the derivative expansion.

From the EFT viewpoint the functions $m^2$, $Z$, $A$ and $B$ are completely unspecified. As anticipated in the Introduction, in order to estimate them we consider a particular toy example for the microscopic interaction of the dark matter particle with the superfluid:
\begin{align}
S_\text{dark}&=-\int d^4x\bigg[ |\partial\chi|^2+m_\chi^2|\chi|^2 +\frac{1}{2}(\partial\phi)^2+\frac{m_\phi^2}{2}\phi^2 \notag \\
&\qquad \qquad \;\; \quad +g_\chi m_\chi \phi |\chi|^2+g_\text{He}\phi\,n\bigg]\,,
\end{align}
where $n$ is the helium number density. Note that, in general, the mediator $\phi$ might couple to any mesoscopic scalar operator $\mathcal{O}$ of the superfluid which play the role of an order parameter. In general, this operator will be originated in the UV from a coupling between the dark sector and the Standard Model as, for example, a coupling between $\phi$ and the quark field. A detailed knowledge of the structure of $\mathcal{O}$ can be obtained, for example, via Monte Carlo methods~\cite{Andreoli:2018etf}. If one neglects spin-dependent couplings, which are expected to be suppressed by the nucleon mass, dimensional analysis tells us that $\mathcal{O}\sim n$, and for the sake of the present work, and in absence of a specific model, it is sufficient to choose $\mathcal{O}=n$. 
Note that while the coupling $g_\text{He}$ between the dark sector and the helium is necessarily small, the smallness of $g_\chi$ is an assumption of our model. We can now match the action~\eqref{eq:Seff1} with the one above and extract the unknown couplings.

The scalar coupling between $\phi$ and $n$ cannot generate effective operators with spin different from zero. It then readily follows that, for the theory under consideration, $A(X)=B(X)=0$. 

\begin{figure}[t]
\centering
\includegraphics[width=0.35\textwidth]{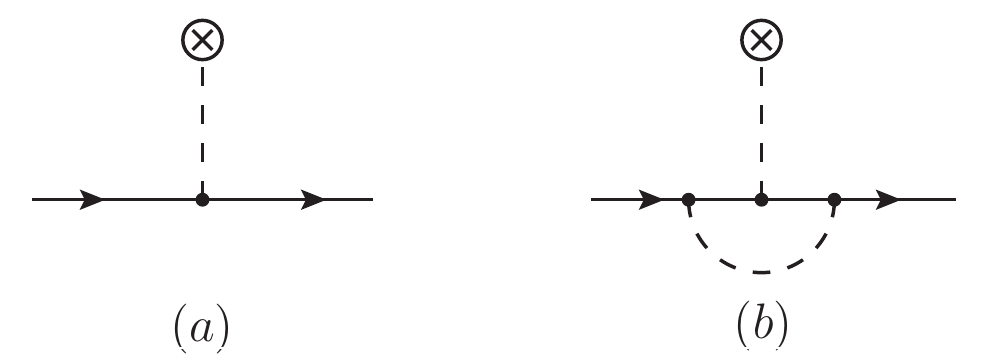}
\caption{Leading and next-to-leading order corrections to the dark matter propagator on the superfluid background. The crossed circle represents the vev of the superfluid operator $\mathcal{O}$.} \label{fig:propagator}
\end{figure}

On the superfluid background, the number density acquires a vev, $\bar n$, which induces a tadople for the mediator, which can then modify the dark matter propagator through the processes reported in Figure~\ref{fig:propagator}. At lowest order in $g_\chi$ and $g_\text{He}$, this induces a shift in the dark matter mass given by
\begin{align}
m^2(\mu) = m_\chi^2-g_\chi g_\text{He}\frac{m_\chi}{m_\phi^2}\bar n(\mu)\,,
\end{align}
where $\bar n$ is a function of the chemical potential on the background. Corrections to the dark matter wave function are only generated by higher order diagrams like the one in Figure~\ref{fig:propagator}b, which we neglect. Hence $Z=1$ at lowest order\footnote{It is likely that the $Z(X)$ coupling will be suppressed anyway due to the nonrelativistic nature of the dark matter. We are grateful to Riccardo Penco for pointing this out.}.

Now that we have estimated the effective mass of the dark matter in the superfluid, the action describing the interaction with the phonon is easily found (in terms of canonical fields) expanding Eq.~\eqref{eq:Seff1}, with  $A=B=0$, for small fluctuations around equilibrium:
\begin{align} \label{eq:Seff}
\begin{split}
S_\text{eff}\supset&\int d^4x\bigg[ -g_1\sqrt{\frac{\mu}{\bar n}}c_s\dot\pi + \frac{g_1}{2}\frac{c_s^2}{\bar n}(\bm \nabla\pi)^2 \\
&\qquad\quad \;\,- \frac{g_2}{2}\frac{\mu c_s^2}{\bar n}\dot\pi^2\bigg]|\chi|^2\,,
\end{split}
\end{align}
with the effective couplings being
\begin{align} \label{eq:gn}
g_n=\frac{d^n m^2}{d\mu^n}=-g_\chi g_\text{He}\frac{m_\chi}{m_\phi^2}\frac{d^n\bar n}{d\mu^n}\,.
\end{align}
From Eqs.~\eqref{eq:Sbulk}~and~\eqref{eq:Seff} we can then find the Feynman rules for the dark matter-phonon(s) vertex and the phonon self-interaction:
\begin{subequations}
\begin{align}
\includegraphics[width=0.12\textwidth,valign=c]{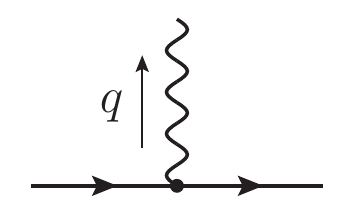} \!\!&= g_1\sqrt{\frac{\mu}{\bar n}}c_s\omega\,, \label{eq:rule1} \\
\includegraphics[width=0.14\textwidth,valign=c]{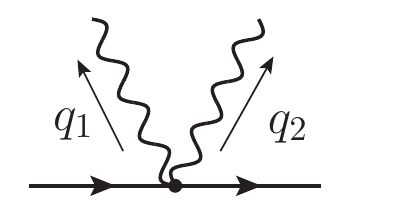} \!\!\!\!\!\!\!\!&= i\frac{c_s^2}{\bar n}\left(\mu g_2\omega_{1}\omega_{2} - g_1\bm q_1\cdot\bm q_2\right) \,, \label{eq:rule2} \\
\includegraphics[width=0.14\textwidth,valign=c]{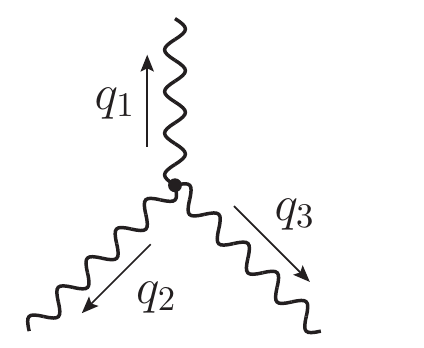}\!\!\!\!\!\!\!\!&=2\sqrt{\frac{\mu}{\bar n}}c_s\bigg[\lambda_3\big(\omega_{1}\bm q_2\cdot\bm q_3+ \omega_{2}\bm q_1\cdot\bm q_3 \notag \\[-1.5em] 
&\quad+\omega_{3}\bm q_1\cdot\bm q_2\big)+3\lambda_3^\prime\omega_{1}\omega_{2}\omega_{3}\bigg] \,. \label{eq:rule3}
\end{align}
\end{subequations}

Let us stress that the discussion above is completely general, true for any relativistic s-wave superfluid at zero temperature. Moreover, the three-phonon vertex (or any other vertex) is obtained straightforwardly, in contrast with  standard techniques~\cite{family1975application}, and it is uniquely determined by the symmetries. In the next section we specify to the case of $^4$He, and work in the nonrelativistic limit, for which $c_s\ll 1$ and $\mu\simeq m_\text{He}$.


\section{Results}

\subsection{One-phonon emission}

When a single phonon is emitted, its energy is not enough to be detected using calorimetric techniques, which have a sensitivity of (at best) 1 meV~\cite{Hochberg:2015pha}. However, it can travel ballistically through the medium and bounce off the walls of the superfluid container until it reaches the surface. It can then induce the evaporation of a helium atom, which could  eventually be observed~\cite{Maris:2017xvi,Hertel:2018aal}. In order for this to happen, the phonon must overcome the surface binding energy of the atom to the rest of the superfluid, which is $\omega_\text{min}=0.62$~meV. Note that this energy range is such that the stability of the phonon against decay is ensured\footnote{We are grateful to D.~McKinsey for pointing this out to us.}~\cite{Maris:1977zz}.

For $^4$He the maximum energy of a phonon is roughly 1 meV. Above that, the dispersion relation ceases to be linear and the collective excitations cannot be described in terms of a phonon degree of freedom. From the EFT viewpoint this means that higher derivative corrections become relevant, and the action~\eqref{eq:Sbulk1} should be supplemented with higher dimensional operators, hence largely losing its predictive power.

Consider the emission of a single phonon. Its maximum energy is $2c_sm_\chi v_\chi$. Since in order for it to be detected it must be $\omega\gtrsim0.62$~meV, and the dark matter velocity is $v_\chi\sim10^{-3}$, it follows that this channel is only effective if $m_\chi\gtrsim1$~MeV.
Given the rule~\eqref{eq:rule1} one finds the emission rate as
\begin{align}
 \frac{d\Gamma}{d\Omega d\omega}=\frac{g_1^2}{32\pi^2}\frac{m_\text{He}\,\omega^2}{v_\chi m_\chi^2\bar n}\delta\left( \cos\theta-\frac{c_s}{v_\chi}-\frac{q}{2m_\chi v_\chi} \right)\,.
\end{align}
As anticipated in the Introduction, energy and momentum conservation force the phonon to be emitted at a specific angle which depends on the momentum of the outgoing phonon, i.e. the \v{C}herenkov angle. Note that the condition that the $\delta$-function has nonzero support, tells us that one cannot emit a phonon with momentum larger than $q_\text{max}=2m_\chi(v_\chi-c_s)$.

Using Eq.~\eqref{eq:gn} together with the thermodynamic identities $dP=\bar n d\mu$ and $dP/d\bar n=m_\text{He}c_s^2$, we can write the effective dark matter-phonon coupling as
\begin{align} \label{eq:g1}
 g_1= -g_\chi g_\text{He}\frac{m_\chi}{m_\phi^2}\frac{\bar n}{m_\text{He}c_s^2}\,.
\end{align}
We then find the rate per unit phonon energy to be
\begin{align}
\frac{d\Gamma}{d\omega}=\frac{g_\chi^2 g_\text{He}^2}{16\pi\, m_\phi^4}\frac{\bar n}{m_\text{He}c_s^4 v_\chi}\omega^2\,.
\end{align}

Since the energy deposited in the superfluid by this process is too small ($\omega\lesssim 1$ meV), it can only be detected via the quantum evaporation. The detection rate per unit target mass is then obtained counting the number of events for which the phonon's energy is in the correct range
\begin{align} \label{eq:N}
N=\int dv_\chi f_\text{MB}(v_\chi)\frac{\rho_\chi}{m_\text{He}\bar n m_\chi}\int_{\omega_\text{min}}^{\omega_\text{max}}d\omega\frac{d\Gamma}{d\omega}\,.
\end{align}
Here $\omega_\text{max}=\text{min}(2c_sm_\chi (v_\chi-c_s), 1\text{ meV})$ is the maximum phonon energy, set by either the momentum of the dark matter times the speed of sound or by the cutoff of the EFT. The local dark matter mass density is $\rho_\chi\simeq0.3$~GeV/cm$^3$~\cite{Bovy:2012tw}, while the helium number density and sound speed at zero temperature are $\bar n\simeq8.5\times10^{22}$~cm$^{-3}$ and $c_s\simeq8.2\times10^{-7}$~\cite{abraham1970velocity}.
Finally the dark matter Maxwell-Boltzmann distribtion in the Milky Way halo is given by
\begin{align}
f_\text{MB}(v_\chi)=4\frac{v_\chi^2}{v_0^2}\frac{e^{-v_\chi^2/v_0^2}\Theta(v_\text{esc}-v_\chi)}{\sqrt{\pi}v_0\text{erf}\left(\frac{v_\text{esc}}{v_0}\right)-2e^{-v_\text{esc}^2/v_0^2}v_\text{esc}}\,,
\end{align}
with $v_0\simeq220$~km/s and $v_\text{esc}\simeq550$~km/s~\cite{LEWIN199687}.

Noticing that the effective coupling for a massive mediator is roughly $g_\chi g_\text{He}/m_\phi^2$, we can estimate the dark matter-helium cross section as
\begin{align}
 \sigma_\text{He}\sim \frac{g_\chi^2 g_\text{He}^2}{16\pi\,m_\phi^4}\frac{m_\chi^2m_\text{He}^2}{(m_\chi+m_\text{He})^2}\simeq A^2 \sigma_{p}\,,
\end{align}
where $\sigma_{p}$ is the dark matter-proton scattering cross section, and $A=4$ for $^4$He. The combination $g_\chi g_\text{He}/m_\phi^2$ can then be expressed in terms of $\sigma_p$ only.

\subsection{Two-phonon emission}

Let us now turn to the process of emission of two phonons by the passing dark matter. Using simple kinematics the authors of~\cite{Schutz:2016tid,Knapen:2016cue} claim that the configuration where the two emitted phonons are back-to-back allows to maximize the energy released to the superfluid, potentially allowing for the detection.

\begin{figure}[t]
 \centering
 \includegraphics[width=0.38\textwidth]{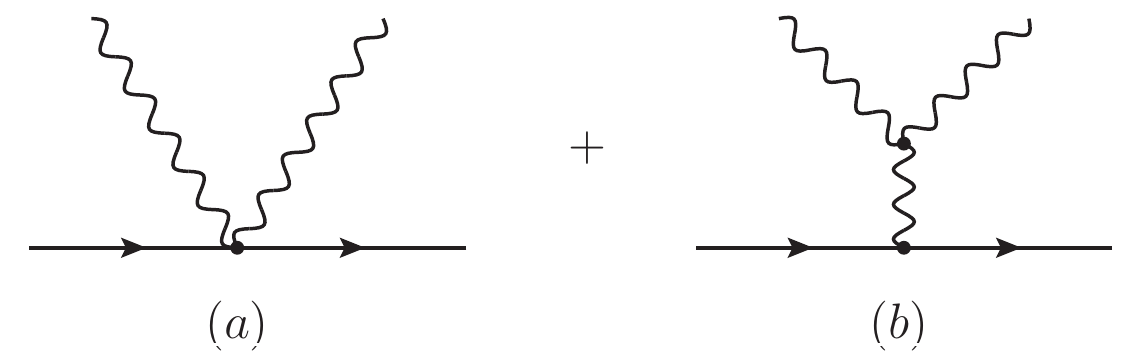}
 \caption{Leading diagrams contributing to the two-phonon emission process.} \label{fig:2phonons}
\end{figure}

Here we re-evaluate it using our EFT. At leading order in $g_\text{He}$, the two diagrams contributing to this process are the ones reported in Figure~\ref{fig:2phonons}. Note that the first one did not appear explicitly before~\cite{Schutz:2016tid,Knapen:2016cue}.  When the two phonons are almost back-to-back, both matrix elements are equally relevant to the process under consideration. One can indeed estimate them with simple dimensional analysis. Reinstating the speed of light, one finds that, with our normalization, $P^\prime=\bar n\sim r_B^{-3}$. Moreover, every derivative of the pressure scales as $m_\text{He}r_B^2$, and $c_s\sim m_\text{He}^{-1}r_B^{-1}$. Given this, one finds, for example, that $\lambda_3^\prime\sim m_\text{He}c_s^2 P^{\prime\prime\prime}/\bar n\sim m_\text{He}r_B^2$. Following these lines, we deduce that both matrix elements are roughly
 \begin{align}
  \mathcal{M}&\sim g_\chi g_\text{He}\frac{m_\chi}{m_\phi^2} \cdot m_\text{He}r_B^2 \cdot \omega_{1}\omega_{2} f(\theta_{12},\omega_{1}/\omega_{2})\,,
 \end{align}
where $\theta_{12}$ is the relative angle between the two outgoing phonons and $f$ an adimensional function, different for the two diagrams. Hence, barring particular kinematical configurations, the two amplitudes are of similar magnitude, as it is also verified numerically. Yet another advantage of the EFT approach is to make manifest the presence of the first diagram. Importantly, the two diagrams turn out to interfere \emph{destructively}.

The effective couplings for the process under consideration can be written in the nonrelativistic limit as
\begin{subequations}
\begin{gather}
 \lambda_3\simeq-\frac{1}{2m_\text{He}}\,,\quad\lambda_3^\prime\simeq\frac{1}{6m_\text{He}c_s^2}-\frac{\bar n}{3c_s}\frac{dc_s}{dP}\,, \\
 g_2\simeq-g_\chi g_\text{He}\frac{m_\chi}{m_\phi^2}\bigg( \frac{\bar n}{m_\text{He}^2c_s^4} - \frac{2\bar n^2}{m_\text{He}c_s^3}\frac{dc_s}{dP} \bigg)\,,
\end{gather}
\end{subequations}
together with the coupling $g_1$ already estimated in Eq.~\eqref{eq:g1}. The derivatives of the sound speed as a function of pressure are extracted from  data~\cite{abraham1970velocity}. Here we consider reference values at atmospheric pressure, for which $dc_s/dP\simeq 8$ m/s/atm. In Table~\ref{tab:couplings} we summarise the numerical values of the effective couplings and parameters in units of energy.

\begin{table}[t]
\centering
\begin{tabular}{c|c||c|c}
\hline\hline
$\bar n$ & $0.65$ keV$^{3}$ & $\lambda_3$ & $-1.3\times10^{-7}$ keV$^{-1}$ \\
\hline
$c_s$ & $8.2\times10^{-7}$ & $\lambda_3^\prime$ & $-8.5\times10^5$ keV$^{-1}$ \\
\hline
$\;d\bar n/d\mu\;$ & $2.7\times 10^{5}$ keV$^2$ & $\;d^2\bar n/d\mu^2\,$ & $-1.4 \times 10^{12}$ keV \\
\hline\hline
\end{tabular}
\caption{Summary of the couplings and parameters extracted from data. The derivatives of the density with respect to the chemical potential can be reduced to derivatives of the sound speed with respect to pressure using standard thermodynamical identities.} \label{tab:couplings}
\end{table}

We now need to evaluate the rate for the emission of two phonons. In a standard Lorentz invariant framework one would boost the system to the center-of-mass of the initial particle, where the computation is simpler, and then boost back to the lab frame. Here, the presence of the medium breaks boost invariance, and the rate must be computed directly in the lab frame.

The final state contains the scattered dark matter particle with momentum $\bm{P}^\prime$ and energy $E^\prime$, and the two phonons with momenta $\bm q_1$ and $\bm q_2$, and energies $\omega_1$ and $\omega_2$. Let $\theta_2$ be the angle of one of the two phonons with respect to the direction of the incoming dark matter particle, $\bm P$. Let $\theta_{12}$ be the angle between the phonons in the final state.
The two-phonon rate is given by
\begin{align}\label{eq:2phonrate}
\Gamma=\frac{1}{8(2\pi)^4c_s^5E}\!\int_{{\cal R}}\!\!d\theta_{12}d\theta_2d\omega_1d\omega_2\, \frac{\omega_2}{P}\frac{|{\cal M}|^2}{\sqrt{1-{\cal A}^2}}\,,
\end{align}
where the matrix element ${\cal M}$ is obtained by the sum of the two diagrams in Figure~\ref{fig:2phonons}, $E$ is the energy of the incoming dark matter, and $\mathcal{R}$ is a suitable integration region --- see below.
The angle $\theta_1$ between $\bm P$ and $\bm q_1$, is given by 
\begin{align}
\cos\theta_1=\cos\theta_{12}\cos\theta_2-{\cal A}\sin\theta_{12}\sin\theta_2
\end{align}
with ${\cal A}=\cos(\phi_{12}-\phi_2)$, where $\phi_{12}$ is the azimuthal angle of $\bm q_1$ in a frame in which $\bm q_2$ is along the $z$-axis ($\theta_{12}$ is the zenith angle), whereas $\phi_2$ is the azimuthal angle of $\bm q_2$ in a frame in which $\bm P$ is along the $z$-axis.   

The momentum delta-function has been integrated over $d^3P^\prime$ leaving 
\begin{align}
d^3q_1 \, d^3q_2 = q_1^2 dq_1 d\phi_{12} d\cos\theta_{12}\; q_2^2 dq_2 d\phi_2 d\cos\theta_2\,.
\end{align}
The energy delta-function has instead been integrated over $\phi_{12}$ to obtain the expression for the phase space in~\eqref{eq:2phonrate}, including the Jacobian
\begin{align}
J=\frac{c_s E^\prime}{\omega_1\sin\theta_{12}\sin\theta_2P}\,.
\end{align}
It follows that the integration region $\mathcal{R}$ is the one over which the delta-function has support. This is defined by those values of $\theta_{12},\theta_{2},\omega_1$ and $\omega_2$ satisfying 
\begin{align} \label{eq:Acondition}
{\cal R}: -1\leq {\cal A}(\theta_{12},\theta_2,\omega_1,\omega_2)\leq +1\,,
\end{align}
where 
\begin{align} \label{eq:aa}
\begin{split}
&{\cal A}(\theta_{12},\theta_2,\omega_1,\omega_2)= \frac{1}{\sin\theta_{12}\sin\theta_2}\Big(\cos\theta_{12}\cos\theta_2 \\ 
&\qquad+\frac{\omega_2}{\omega_1}\cos\theta_2-\frac{\omega_2}{c_s P}\cos\theta_{12}-\frac{\omega_1^2+\omega_2^2}{2\omega_1 c_s P}\,\Big)\,.
\end{split}
\end{align}

The calculation of the integral is conveniently done using Monte Carlo techniques (in particular we took advantage of the Vegas algorithm available in the CUBA library~\cite{Hahn:2004fe}). 

Given the above setup, we impose a number of kinematical cuts to reflect both consistency with the regime of applicability of the EFT as well as experimental constraints. First of all, as in the previous section, we integrate the phonon's momenta only up to $q_\text{max}=1$~keV. Secondly, we require that the momentum flowing in the phonon propagator does not exceed the cutoff of the EFT\footnote{For dimensional reasons the energy and momentum cutoffs are related by $\omega_\text{max}\sim c_s q_\text{max}$.}. This is done imposing that the momenta satisfy
\begin{align}
|\vec q_1+\vec q_2|\leq 1 \text{ keV}\,.
\end{align}
This cut also cures the collinear divergence coming from the propagator at $\theta_{12}=0$.

As explained in the Introduction, there are two possible ways to detect the event. Either the phonons have separately enough energy to induce quantum evaporation on the surface of the superfluid, or the net energy released to the detector is enough to be observed using, say, a Transition Edge Sensor, which we assume to be in thermal equilibrium with the helium bath. We treat them as two independent signatures\footnote{It is possible to envision a setup where these two signatures combined work as a trigger to discriminate the emission of one phonon from that of two phonons.} and, based on which one we are considering, we impose additional cuts. In particular, in the first case we require that both phonons satisfy $\omega_i\geq0.62$ meV, while in the second case we require that $\omega_1+\omega_2\geq 1$ meV. In the following we refer to these alternatives respectively as ``evaporation'' and ``energy deposit''. The total rate per unit time and detector's mass is computed integrating Eq.~\eqref{eq:2phonrate} over the Maxwell-Boltzman distribution, as in Eq.~\eqref{eq:N}.

\begin{figure}[t]
\centering
\includegraphics[width=0.47 \textwidth]{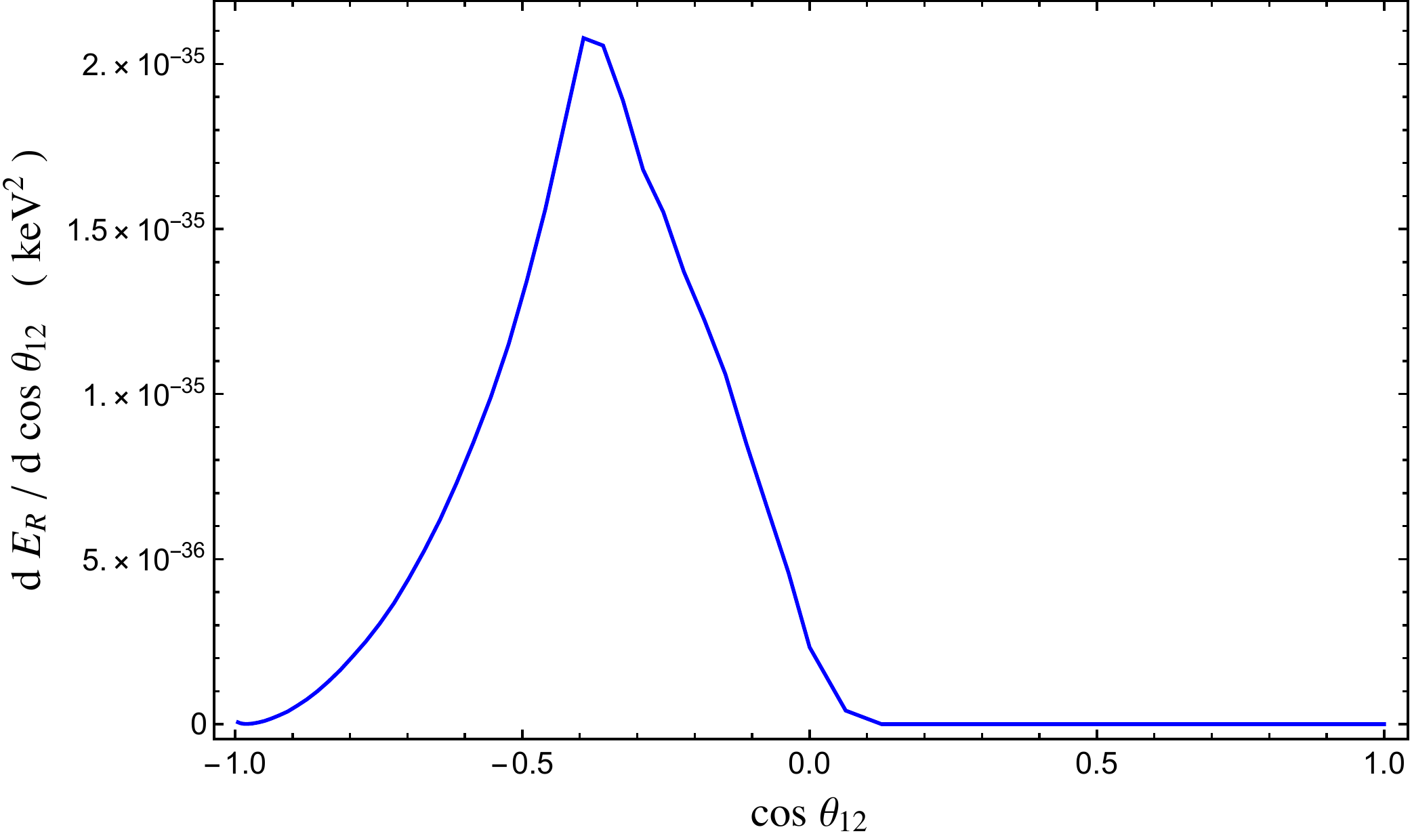}
\caption{Differential distribution of the released energy per unit time as a function of the relative angle between the outgoing phonons. This sample plot is obtained for a dark matter with mass and momentum $m_\chi = 1$ MeV and $P = 1$ keV, and for a dark matter-proton cross section $\sigma_p=10^{-40}$ cm$^2$. We assumed the detection happens via evaporation.} \label{fig:c23}
\end{figure}

In Figure~\ref{fig:c23} we report a sample distribution for the net energy released to the superfluid, $E_R=\omega_1+\omega_2$, per unit time as a function of the relative angle between the two phonons. As one can see, the maximum energy is released when the two phonons are almost back-to-back, although the peak is substantially shifted from $\theta_{12}=\pi$. Such a shift is due to the fact that the $\theta_{12}=\pi$ configuration is forbidden by phase space. In fact, when $\theta_{12}\to\pi$ then $\mathcal{A}\to\infty$, except for the zero measure set where $\omega_1=\omega_2$, and the condition~\eqref{eq:Acondition} is never satisfied.

\subsection{Exclusion region}

\begin{figure}[t]
 \centering
 \includegraphics[width=0.47\textwidth]{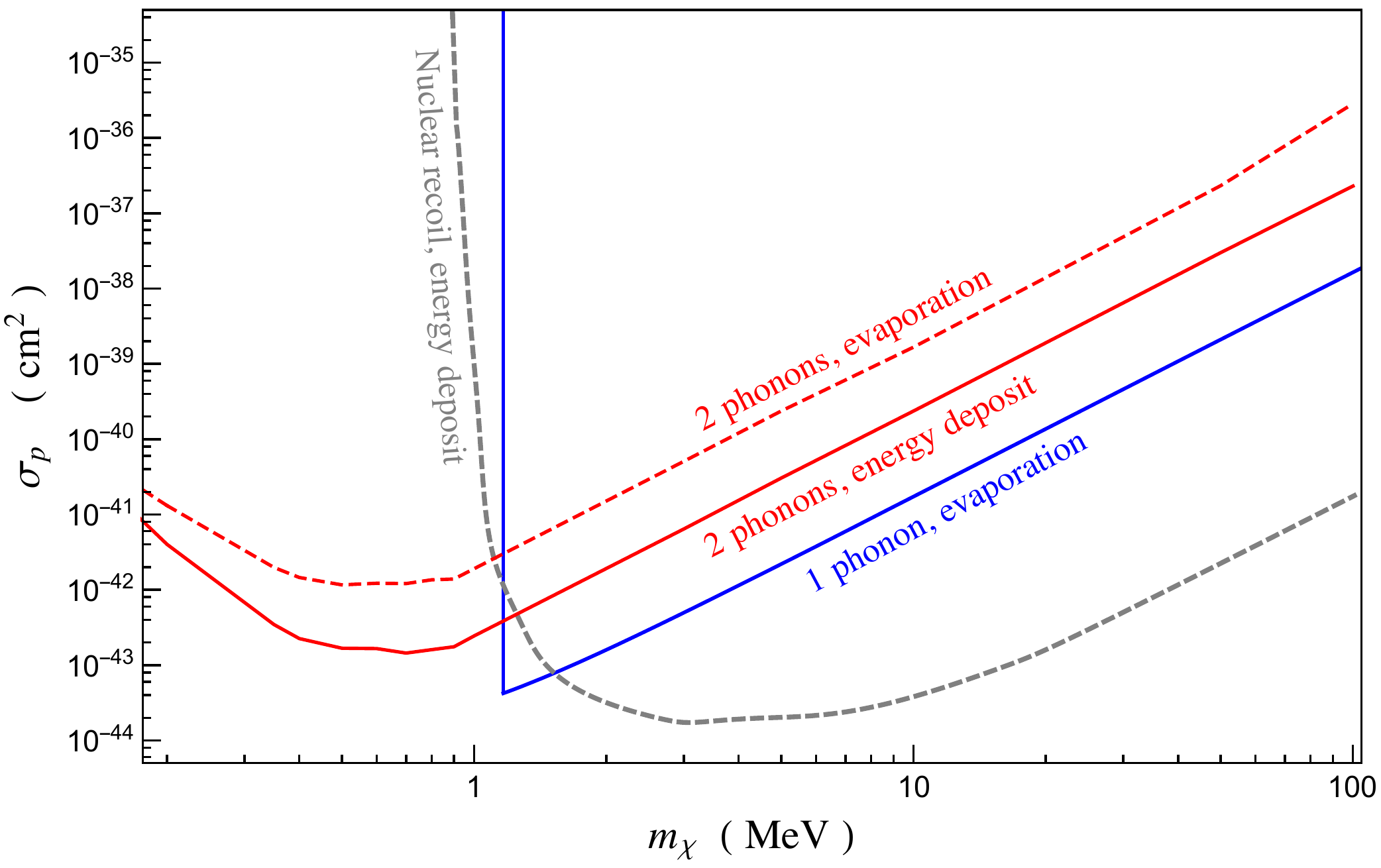}
 \caption{Exclusion region as refered to 95\% C.L., corresponding to 3 events$/\text{kg}/\text{year}$, assuming zero background. The data for nuclear recoil have been taken from~\cite{Knapen:2016cue}. The sharp vertical line in the one-phonon case corresponds to the value of the dark matter mass for which it becomes too light to produce a detectable phonon. The two-phonon emission process remains effective also at masses lighter than 1 MeV.} \label{fig:exclusion1}
\end{figure}

Our predicted exclusion region is reported in Figure~\ref{fig:exclusion1}. We have assumed no background and the sensitivity necessary to detect a net energy deposit of 1 meV, as well as to observe single phonons through quantum evaporation.

As one can see, a combination of observables allows to cover different orders of magnitude for the dark matter mass. For masses below the MeV the two-phonon process is the only one that has the right kinematics to be potentially observed, and it can be relevant for substantially lighter masses. Recall that here we only account for the phononic excitations described by our EFT. The inclusion of higher momentum excitations, like maxons or rotons, opens up a large portion of phase space. Nevertheless, our framework allows to compute the rate for this process up to arbitrarily high masses, beyond the limitations of standard techniques. There are, in fact, no available information on the helium dynamical structure function for this kinematical regime~\cite{Schutz:2016tid}. 
Above 1 MeV there are two dominant processes: hard nuclear recoil and the emission of a single phonon. The first one is a process where the dark matter energy is released to short wavelength modes rather than collective, long wavelength excitations and can then only be detected via energy deposit. The second one, although less effective, can be detected via quantum evaporation and offers a valuable independent channel, relevant for a different range of exchanged momentum.

\section{Conclusion}

In this work we explored a new approach to the problem of the search for sub-GeV dark matter using superfluid $^4$He. From the EFT viewpoint, the interaction between the dark matter and the superfluid phonon is easily described in a quantum field theory language. This allowed us to perform a number of improvements with respect to previous studies (e.g. to formulate the problem in a quantum field theory language, and to reach higher values of the dark matter mass), as well as to have easy access to all sorts of differential distributions, which are crucial for experimental analyses and were not available before. The current EFT approach is however only valid for the description of phonons and does not incorporate higher momentum excitations.

With these information at hand, one can start envisioning different experimental devices that take advantage of the event distributions. For example, given that the emission of a single phonon happens at a fixed angle with respect to the direction of the dark matter, one could think about possible designs for directional detectors.

Other signatures of the interaction with dark matter can involve different excitations of $^4$He, like quantized vortices and rotons. In particular, the latter ones probably contribute to a large portion of the available phase space for the processes considered in this work. An effective theory for the description of superfluid vortices has been developed in~\cite{Horn:2015zna}, while the first important steps towards the development of a field theory for the description of rotons have been made in~\cite{Nicolis:2017eqo}.

The EFT we presented here is valid for the ideal case of a zero-temperature superfluid. It would be interesting to study the effects that finite temperature has on the observable we considered here, especially on the phonon's lifetime~\cite{Hertel:2018aal}. Away from the zero-temperature limit a superfluid presents two different kinds of excitations: the standard superfluid phonon and the phonons of an ordinary fluid --- the so-called two-fluid model~\cite{landau2013fluid}. The possible interactions between the two could be relevant for our analysis. An EFT for the description of a finite temperature superfluid has been developed in~\cite{Nicolis:2011cs}, although its quantization is a nontrivial task (see e.g.~\cite{Endlich:2010hf,Crossley:2015evo,Glorioso:2017fpd}).

Lastly, the only input necessary to our analysis are the symmetries of the dark sector and its coupling to ordinary matter. It can then be extended to different models of dark matter. The exclusion plots and distibutions shown were determined in a dark matter toy model. We leave these numerous possible upgrades for future work.

\vspace{1em}

\acknowledgments{We are grateful to G.~Cavoto, T.~Lin, R.~Penco, F.~Piccinini, R.~Rattazzi, L.~Vecchi and S.~Will for helpful discussions, and to A.~Nicolis for collaboration in the early stages of this project. The work done by A.E. is supported by the Swiss National Science Foundation under contract 200020-169696 and through the National Center of Competence in Research SwissMAP.}

\bibliographystyle{apsrev4-1}
\bibliography{biblio}

\end{document}